\documentclass[reprint,prl,twocolumn,footinbib,aps,nobalancelastpage]{revtex4-2}
\usepackage{amsmath}
\usepackage{amsfonts}
\usepackage{amssymb}
\usepackage{mathtools}
\usepackage{graphicx}
\usepackage{xcolor}
\usepackage[colorlinks=True,citecolor=gray,linkcolor=black,urlcolor=gray]{hyperref}
\usepackage{hypcap}
\usepackage{braket}
\usepackage{yfonts}
\usepackage{bm}
\usepackage[mathscr]{eucal}
\newcommand\eq[1]{\begin{align}#1\end{align}}
\newcommand\ia{{i_A}}
\newcommand\ja{{j_A}}
\newcommand\ib{{i_B}}
\newcommand\jb{{j_B}}
\newcommand\iapp{{i_{A^{\prime\prime}}}}
\newcommand\japp{{j_{A^{\prime\prime}}}}
\newcommand\ibpp{{i_{B^{\prime\prime}}}}
\newcommand\jbpp{{j_{B^{\prime\prime}}}}
\newcommand\pu{\mathcal{P}}
\newcommand\nh{N_\mathcal{H}}

\definecolor{myBlue}{RGB}{31,119,180}
\definecolor{myOrange}{RGB}{255,127,14}
\definecolor{myGreen}{RGB}{44,160,44}
\definecolor{myRed}{RGB}{214,39,40}
\definecolor{myPurple}{RGB}{148,103,189}

\makeatletter
\def\p@figure{\color{myBlue}}
\def\p@equation{\color{myRed}}
\makeatother

\begin{document}
\title{Hilbert-space correlations beyond multifractality and bipartite entanglement in many-body localised systems}
\author{Sthitadhi Roy}
\email{sthitadhi.roy@icts.res.in}
\affiliation{International Centre for Theoretical Sciences, Tata Institute of Fundamental Research, Bengaluru 560089, India}

\begin{abstract}
Eigenstates of many-body localised (MBL) systems are characterised by area-law bipartite entanglement along with multifractal statistics of their amplitudes on Hilbert space. At the same time, sparse random pure states with fractal statistics are not compatible with area-law and necessarily exhibit volume-law entanglement. This raises the question that what correlation functions of Hilbert-space amplitudes MBL eigenstates must possess for their area law entanglement to be compatible with their multifractality. In this work, we identify and compute such appropriate Hilbert-space correlations which carry information of entanglement. We find that, for MBL eigenstates, these correlations are qualitatively different not only from those of ergodic states but also of sparse random states with fractal statistics. This enables us to show that indeed the said correlations lie at the heart of the coexistence of area-law entanglement and multifractality for MBL eigenstates.
\end{abstract}

\maketitle

Sufficiently strong quenched disorder in an isolated quantum many-body system can drive it into a \emph{many-body localised} (MBL) phase through a dynamical phase transition, at least in one spatial dimension~\cite{gornyi2005interacting,basko2006metal,oganesyan2007localisation,pal2010many,imbrie2016many,nandkishore2015many,abanin2017recent,alet2018many,abanin2019colloquium}. Such a system, unlike a chaotic system, fails to thermalise under its own dynamics. The many-body localisation transition is an eigenstate phase transition; across the transition, the properties of individual eigenstates at arbitrary energy densities change sharply. In real space, the bipartite entanglement transitions from a volume law in the ergodic phase to an area law in the MBL phase~\cite{serbyn2013local,bauer2013area,luitz2015many,geraedts2016many,geraedts2017characterising,khemani2017critical}. At the same time, the statistics of eigenstate amplitudes in Hilbert space go from fully ergodic, delocalised to multifractal throughout~\footnote{Of course in the extreme limit of infinite disorder strength, the eigenstates are product states and the fractal dimension is zero.} the MBL phase~\cite{deluca2013ergodicity,mace2019multifractal,detomasi2020rare,roy2021fockspace}. However, the precise connection between these two aspects is unclear and constitutes the central question of this work.

The question is pertinent as multifractality in Hilbert space, in general, does not imply area-law bipartite entanglement. Indeed, it was shown that sparse random pure states with multifractal statistics exhibit volume-law entanglement; the fractal dimension only controls the coefficient of the volume law~\cite{detomasi2020multifractality}. This, therefore, poses a sharp question -- what features of MBL eigenstates allow for multifractality on Hilbert space to coexist with area-law entanglement in real space?

In this work, we show that appropriately defined Hilbert-space correlations encode information about entanglement in real space. As such, they distinguish area-law entangled states not only from volume-law entangled ergodic states but also from volume-law entangled multifractal states. These are multipoint correlations on Hilbert space and hence manifestly go beyond multifractal statistics which formally can be expressed via one-point correlations.

For concreteness, we consider a chain of spins-1/2 (denoted via the set of Pauli matrices $\{\sigma^\mu\}$) of length $L$. We bipartition the system into two equal halves, $A$ and $B$. For a state $\rho = \ket{\psi}\bra{\psi}$, the $n^\mathrm{th}$ R\'enyi entropy of entanglement between the bipartitions is given by $S_n^A = \frac{1}{1-n}\ln\mathrm{Tr}[\rho_A^n]$ with $\rho_A = \mathrm{Tr}_B\rho$. Here we focus solely on $n=2$ as $S^A_2=-\ln\mathrm{Tr}[\rho_A^2]$ where $\pu=\mathrm{Tr}\rho_A^2$ is defined as the \emph{purity} of the state. It is one of the simplest measures of the mixed-ness of the reduced density matrix $\rho_A$. In particular, $\pu$ is expected to be exponentially small $L$ in the ergodic phase and independent of $L$ in the MBL phase. We next discuss how $\pu$ can be expressed in terms of Hilbert-space correlations.

We work in the basis of $\sigma^z$-product states, denoted by $\{\ket{\mathcal{I}}\}$, and decompose each basis state as $\ket{\mathcal{I}}=\ket{\ia}\otimes\ket{\ib}$ where $\ket{i_{A(B)}}$ corresponds to the spin configuration in subsystem $A(B)$ in state $\ket{\mathcal{I}}$. Any state $\ket{\psi}$ can then be decomposed as
\eq{
	\ket{\psi} = \sum_{\ia,\ib}\psi_{\ia\ib}\ket{\ia}\otimes\ket{\ib}\,,
	\label{eq:psidecomp}
}
where $\{\psi_{\ia\ib}\}$ denotes the set of amplitudes on Hilbert space. Using Eq.~\ref{eq:psidecomp}, $\pu$ can be expressed as 
\eq{
	\pu=\sum_{\ia,\ib;\ja,\jb}\psi_{\ia\ib}\psi_{\ia\jb}^\ast\psi_{\ja\ib}^\ast\psi_{\ja\jb}\,,
	\label{eq:purity}
}
which is a sum of four-point correlations between the state amplitudes on Hilbert space. The amplitudes on the four basis states involved in each term of the sum in Eq.~\ref{eq:purity} are made up of two basis states from each subsystem, $(i_{A/B},j_{A/B})$. In order to recast $\pu$ as a spatial correlation on Hilbert space, it is useful to organise the basis states in terms of two Hamming distances, $r_A$ and $r_B$, corresponding to each subsystem; $r_{A(B)}$ is the number of spins in subsystem $A(B)$ different between the two configurations in question. Specifically, between two configurations $\ket{\mathcal{I}}=\ket{\ia}\ket{\ib}$ and $\ket{\mathcal{J}}=\ket{\ja}\ket{\jb}$, the Hamming distance is $r_{\mathcal{I}\mathcal{J}}=r_{\ia\ja}+r_{\ib\jb}$. With this convention for distances, $\pu$ can be written in terms of a Hilbert-spatial correlation, $C(r_A,r_B)$, as $\pu=\sum_{r_A=0}^{L/2}\sum_{r_B=0}^{L/2}C(r_A,r_B)$ where the $C(r_A,r_B)$ is defined as 
\eq{
	C(r_A,r_B) = \sum_{\substack{\ia,\ja:r_{\ia\ja}=r_A \\\ib,\jb: r_{\ib\jb}=r_B}}\psi_{\ia\ib}\psi_{\ia\jb}^\ast\psi_{\ja\ib}^\ast\psi_{\ja\jb}\,.
	\label{eq:CrArB}
}
Note that $C(r_A=0,r_B=0)=\sum_{\ia,\ib}|\psi_{\ia\ib}|^4$ is the inverse participation ratio (IPR) of the state. However, the entanglement properties of the state contained in $\pu$ depends on the entire function $C(r_A,r_B)$. This already indicates that multifractality is not enough to determine entanglement. In the following, we will average the correlation function over disorder realisations and states and denote it by $\braket{C(r_A,r_B)}$. This leads us to the average purity $\braket{\pu}$ and hence to an annealed second R\'enyi entropy $S^A_{2,\mathrm{ann}} = -\ln\braket{\pu}$. However, like the average second R\'enyi entropy, $S^A_{2,\mathrm{ann}}$ is also expected to transition from a volume law in the ergodic phase~\cite{lubkin1978entropy,page1993average,lu2019renyi} to an area law in the MBL phase~\cite{monthus2016manybody}.

\begin{figure}[!t]
\includegraphics[width=0.9\linewidth]{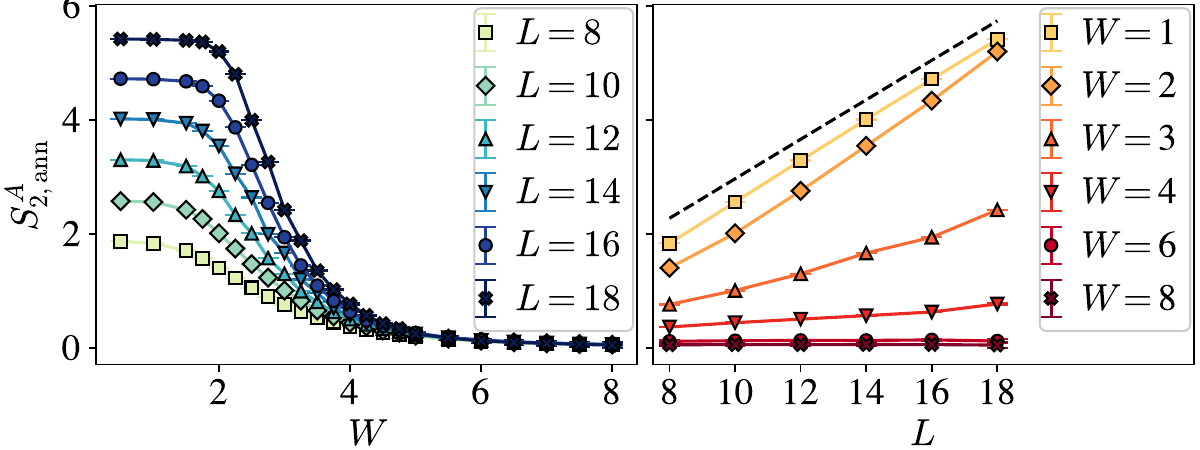}
\caption{The annealed second R\'enyi entropy, $S_{2,\mathrm{ann}}^A=-\ln\braket{\pu}$ as a function of $W$ for different $L$ (left), and as a function of $L$ for different $W$ (right). The black dashed line is $\propto (L/2)\ln2$. The data clearly shows volume-law for $W<W_c$ and area-law for $W>W_c$.}
\label{fig:s2}
\end{figure}

To study how the correlation $C(r_A,r_B)$ changes across the MBL transition, we consider an archetypal model, namely a disordered spin-chain described by the Hamiltonian
\eq{
	H = \sum_{\ell=1}^{L-1}J_\ell\sigma^z_{\ell}\sigma^z_{\ell+1} +\sum_{\ell=1}^L[h_\ell\sigma^z_{\ell} + \Gamma\sigma^x_\ell]\,,
	\label{eq:H}
}
where $J_\ell\in[J-\Delta,J+\Delta]$ and $h_\ell\in[-W,W]$ are independent random numbers drawn from a uniform distribution. We use $J=\Gamma=1$ and $\Delta=0.2$. For these parameters and the kind of system sizes accessible to exact diagonalisation (ED), the model has an estimated critical disorder strength of $W_c\approx 3.75$~\cite{abanin2021distinguishing}. However, recent works have suggested that a genuine MBL phase, stable in the thermodynamic limit, can set in only at much larger values of $W$ for standard disordered models~\cite{suntajs2020quantum,morningstar2021avalanches,sels2022bath} and the apparent localisation for finite systems at $W>W_c$ can be understood partially via many-body resonances~\cite{,gopalakrishnan2015lowfrequency,garratt2021resonances,crowley2022constructive,garratt2022resonant,long2022phenomenology} or alternatively via the proximity of the MBL phase to an Anderson localised one~\cite{vidmar2021phenomenology}. We take the view that the phenomenology observed at $W>W_c$ for system sizes accessible to ED persist in the thermodynamic limit, albeit for much larger $W$. We compute $C(r_A,r_B)$ and $\pu$ using a few eigenstates from the middle of the spectrum extracted from ED and average the data over several thousand samples.

In Fig.~\ref{fig:s2}, we show $S^A_{2,\mathrm{ann}}$ for different $L$ and $W$. The data clearly shows that in the ergodic regime ($W<W_c$), the entanglement follows a volume law, $S^A_{2,\mathrm{ann}}\sim  sL$ with $0<s\le (\ln 2)/2$ whereas in the MBL regime ($W>W_c$), it follows an area-law, $S^A_{2,\mathrm{ann}}\sim\mathcal{O}(1)$. Equivalently, for $W<W_c$, the average purity $\braket{\pu}\sim \nh^{-\gamma}$ where $\nh=2^L$ is the total Hilbert-space dimension and $0<\gamma\le 1/2$ whereas for $W>W_c$, $\braket{\pu}$ is independent of $L$.

\begin{figure}[t]
\includegraphics[width=0.9\linewidth]{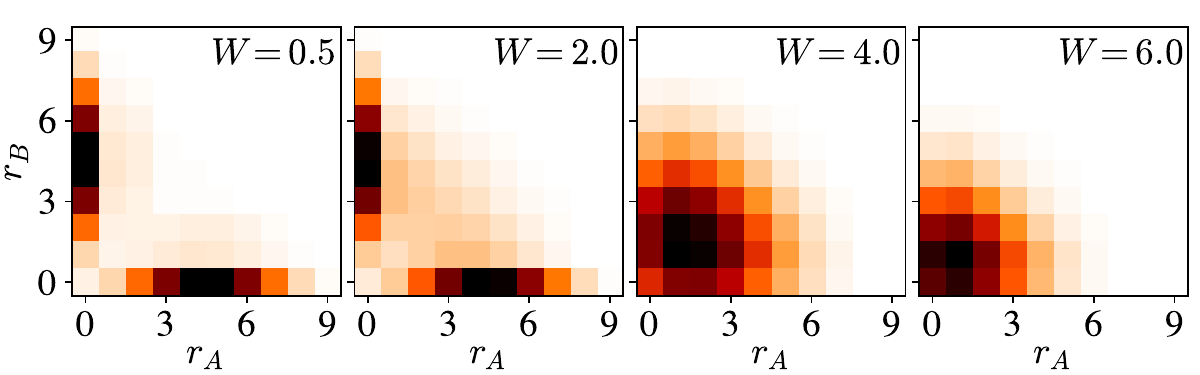}
\caption{The average correlation function, $\braket{C(r_A,r_B)}$ (see Eq.~\ref{eq:CrArB}), as a heatmap in the $(r_A,r_B)$ for different values of $W$. Data for $L=18$ and the colours in each panel are normalised to the respective maximum and minimum values.}
\label{fig:crarb}
\end{figure}

\begin{figure}[!t]
\includegraphics[width=0.9\linewidth]{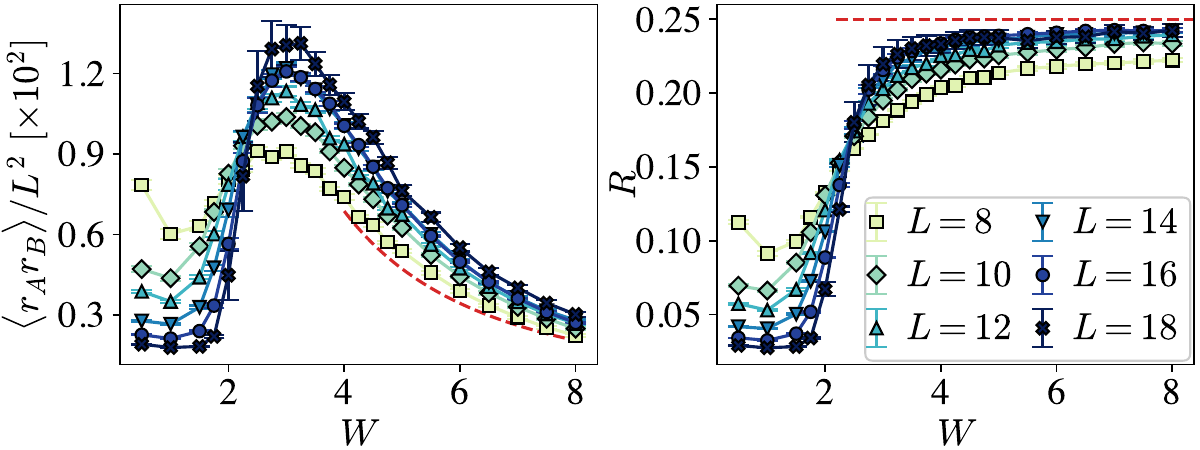}
\caption{The covariance $\braket{r_Ar_B}$ defined in Eq.~\ref{eq:rarab} rescaled with $L^2$ (left) and the ratio of $\braket{r_Ar_B}$ to $\braket{r_A+r_B}^2$ defined in Eq.~\ref{eq:R} (right), as a function of $W$ for different $L$. The red dashed lines in both panels correspond to the case with $J_\ell=0$.}
\label{fig:rarbR}
\end{figure}

Having established the behaviour of $S^A_{2,\mathrm{ann}}$ or equivalently that of $\braket{\pu}$ across the transition, we next turn to the correlation function $C(r_A,r_B)$. The results for it  are presented in Fig.~\ref{fig:crarb}, which clearly shows a stark difference in the profiles of $\braket{C(r_A,r_B)}$ between the ergodic and MBL regimes. In the former, the correlations are totally dominated by $r_A=0$ or $r_B=0$. On increasing $W$ correlations develop for $r_A,r_B\neq 0$ and in the MBL regime, $\braket{C(r_A,r_B)}$ is peaked at some $r_A,r_B\neq 0$ which again shifts towards $r_A,r_B=0$ as $W$ becomes very large. To make this quantitative, note that $\braket{C(r_A,r_B)}$ can be interpreted as an (unnormalised) probability distribution over $r_A$ and $r_B$ and its covariance quantifies precisely the information said in words above. Appropriately normalised, the covariance is defined as 
\eq{
	\braket{r_Ar_B} = \braket{\pu}^{-1}\sum_{r_A=0}^{L/2}\sum_{r_B=0}^{L/2}r_A r_B \braket{C(r_A,r_B)}\,,
	\label{eq:rarab}
}
and the results are shown in Fig.~\ref{fig:rarbR}(left). In the ergodic regime, $\braket{r_Ar_B}/L^2$ decays towards zero with increasing $L$ consistent with the fact that $C(r_A,r_B)$ is finite only for $r_A= 0$ or $r_B= 0$. In the MBL regime on the other hand, $\braket{r_Ar_B}\propto L^2$ with the constant of proportionality decaying monotonically with $W$ as the peak shifts towards the origin. In order to account for this decay one can renormalise $\braket{r_Ar_B}$ by defining~\footnote{see supplementary material \cite{supp} for an alternative renormalisation}
\eq{
	R = \frac{\braket{r_A r_B}}{\braket{r_A+r_B}^2}\,,
	\label{eq:R}
}
where $\braket{r_A+r_B}$ is defined in the same way as in Eq.~\ref{eq:rarab}. The results for $R$ are shown in Fig.~\ref{fig:rarbR}(right), where it is clear that in the ergodic regime $R\to 0$ as $L\to \infty$ whereas in the MBL regime $R$ is finite and $L$-independent, and it approaches $1/4$ in the limit of $W\to\infty$. In fact, from the data in Fig.~\ref{fig:rarbR}, it is tempting to identify $R$ as a diagnostic for the many-body localisation transition.

The phenomenology that emerges from the collection of numerical results above is that the Hilbert-space correlation $C(r_A,r_B)$ which contains information about the purity $\pu$ indeed changes qualitatively across the many-body localisation transition. We next try to understand some of these features analytically.

Deep in the ergodic phase, the eigenstates are very well approximated as random Gaussian vectors with zero mean and standard deviation of $\nh^{-1/2}$~\cite{mehta2014random,haake2010quantum},
\eq{
	\braket{\psi_{\ia\ib}}=0;\quad\braket{\psi_{\ia\ib}\psi_{\ja\jb}}=\nh^{-1}\delta_{\ia\ja}\delta_{\ib\jb}\,.
	\label{eq:randomstate}
}
Note that the model in Eq.~\ref{eq:H} is time-reversal symmetric and hence it is sufficient to consider real eigenstate amplitudes. Using Eq.~\ref{eq:randomstate} in Eq.~\ref{eq:CrArB} and performing the average using Wick's probability theorem, we obtain
\eq{
	\braket{C(r_A,r_B)}=\frac{\delta_{r_A0}\delta_{r_B0}+\binom{L/2}{r_A}\delta_{r_B0}+\binom{L/2}{r_B}\delta_{r_A0}}{\nh}\,,
	\label{eq:crarberg}
}
which is exactly the form seen in Fig.~\ref{fig:crarb} for $W\ll W_c$. A crucial ingredient in the above result is that the eigenstate amplitudes can be considered independent random numbers. Since $\braket{C(r_A,r_B)}$ in Eq.~\ref{eq:crarberg} is vanishing unless $r_A=0$ or $r_B=0$, $\braket{r_Ar_B}$ also vanishes trivially. Summing Eq.~\ref{eq:crarberg} over $r_A$ and $r_B$ we obtain $\braket{\pu} = 2\nh^{-1/2}(1+1/2\nh^{1/2})$ such that $S^A_{2,\mathrm{ann}}\approx (L/2)\ln 2$ for large $L$, thus recovering the volume-law entanglement seen in Fig.~\ref{fig:s2}.

We next discuss the MBL phase. It is expected that deep in this phase, the interacting model is perturbatively connected to its non-interacting limit with $J_\ell=0$~\cite{serbyn2013local,huse2014phenomenology,imbrie2017local}. In this limit, the model is trivially MBL as it is a set of non-interacting spins. The eigenstates are tensor products over the states of the two subsystems $\ket{\psi}=\sum_{\ia}\psi_{\ia}\ket{\ia}\otimes\sum_{\ib}\psi_{\ib}\ket{\ib}$ which trivially implies $\pu=1$. The tensor product structure also leads to  $C(r_A,r_B)$ splitting into a product of two-point correlations for each subsystem as 
\eq{
	C(r_A,r_B) = \sum_{\substack {\ia,\ja:\\r_{\ia\ja}=r_A}}|\psi_{\ia}\psi_{\ja}|^2\sum_{\substack {\ib,\jb:\\r_{\ib\jb}=r_B}}|\psi_{\ib}\psi_{\jb}|^2\,.
	\label{eq:crarbsep}
}
Exploiting the non-interacting nature of the spins, the two-point correlations in Eq.~\ref{eq:crarbsep} can be evaluated exactly~\cite{roy2021fockspace} to give
\eq{
	\braket{C(r_A,r_B)} = \binom{\frac{L}{2}}{r_A}\binom{\frac{L}{2}}{r_B}p^{r_A+r_B}(1-p)^{L-r_A-r_B}\,,
	\label{eq:crarbmbl}
}
where $p=(\tan^{-1}W)/2W$. This immediately implies that $\braket{C(r_A,r_B)}$ is sharply peaked at $(r_A,r_B)=(pL/2,pL/2)$ which is qualitatively different from that of the ergodic regime. The form in Eq.~\ref{eq:crarbmbl} also yields $\braket{r_Ar_B}=p^2L^2/4$ consistent with the $L^2$ scaling of $\braket{r_Ar_B}$ in the MBL phase in the data in Fig.~\ref{fig:rarbR}(left). Additionally, we have $\braket{r_A}=\braket{r_B}=pL/2$ such that $R=1/4$. This is indeed the value that the data in Fig.~\ref{fig:rarbR}(right) approaches for large $W$.

\begin{figure}[!b]
\includegraphics[width=0.9\linewidth]{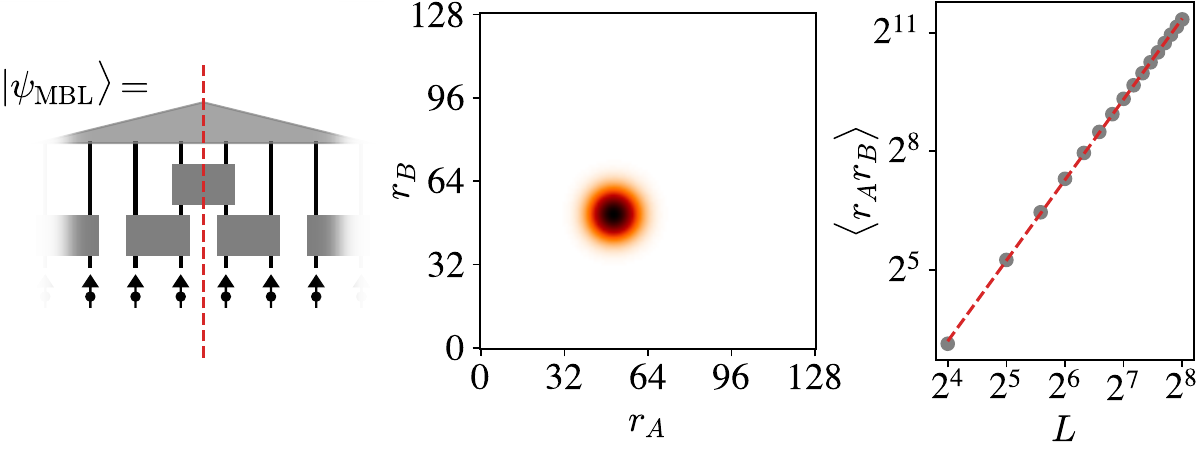}
\caption{Left: The state $\ket{\psi_\mathrm{MBL}}$ is constructed by a finite-depth local unitary made out of Haar random unitaries acting on a product state. We ignore the unitaries in the second layer away from the bipartition as they do not affect the entanglement. Middle: $\braket{C(r_A,r_B)}$ for the state above for $L=256$ as a colourmap. Right: The corresponding $\braket{r_Ar_B}$ as function of $L$ scales as $L^2$ as denoted by the red dashed line.}
\label{fig:haarmbl}
\end{figure}

The above results for non-interacting spins along with the numerical results presented in Figs.~\ref{fig:crarb} and \ref{fig:rarbR} suggest that the features of $\braket{C(r_A,r_B)}$ are robust for generic MBL states. However to make this more concrete, we next present analytical results for an MBL state beyond non-interacting spins. Since the MBL eigenstates are related to product states via finite-depth local unitary operators~\cite{serbyn2013local,huse2014phenomenology,bauer2013area}, we can construct a proxy for an MBL state as $\ket{\psi_\mathrm{MBL}} = U\otimes\prod_{\ell=1}^L\ket{\uparrow}$, where $U$ is the finite-depth local unitary depicted in Fig.~\ref{fig:haarmbl}(left) and we can choose the parent product state to be the all-up state without any loss of generality. Each 2-local gate depicted by the grey rectangles is a $4\times4$ Haar-random unitary. This along with the geometry of $U$ allows us to derive an expression for $\braket{C(r_A,r_B)}$ analytically. The expressions are unwieldy enough that we relegate them to the supplementary material~\cite{supp} and instead show the results graphically in Fig.~\ref{fig:haarmbl} for large $L$. The middle panel again shows that $\braket{C(r_A,r_B)}$ is sharply peaked at a value of $r_A,r_B \sim L$ and correspondingly $\braket{r_Ar_B}\sim L^2$ as shown in the right panel.

The conclusion that one draws from all of the above is that the Hilbert-space correlation function $\braket{C(r_A,r_B)}$ which contains information about $\braket{\pu}$ is qualitatively different in the ergodic and MBL phases and is intimately related to the volume-law and area-law bipartite entanglement in the two phases respectively. In order to show that these correlations are indeed at the heart of area-law entanglement in the MBL phase we next discuss the fate of these correlations for sparse random pure states (SRPS) which exhibit volume-law bipartite entanglement along with (multi)fractality on Hilbert space.

\begin{figure}[!t]
\includegraphics[width=0.9\linewidth]{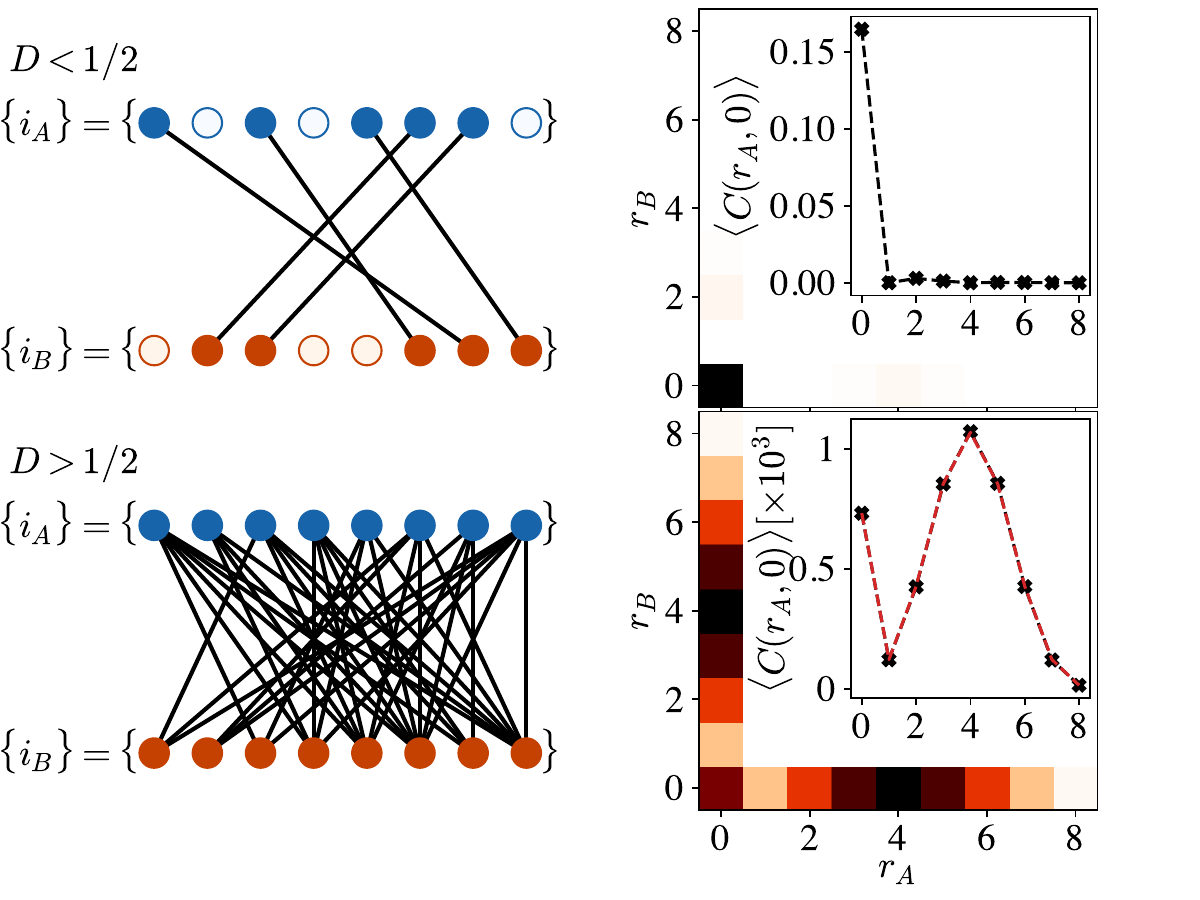}
\caption{Left: Interpreting the SRPS geometrically as links between $\ia$ and $\ib$ for every non-zero $\psi_{\ia\ib}$. The filled circles denote $i_A$ and $i_B$ which have at least one link connected to them whereas the empty circles denote those that have no links.  For $D\lessgtr 1/2$, the link terminations sample the sets $\{\ia\}$ and $\{\ib\}$ sparsely/densely. Right: Numerical results for $\braket{C(r_A,r_B)}$ for SRPS with $D=0.25$ (top) and $D=0.75$ (bottom). The insets show $\braket{C(r_A,0)}$ (errorbars are smaller than marker sizes) with the red dashed curve (bottom) denoting the result in Eq.~\ref{eq:sprsDg} for $D>1/2$.}
\label{fig:sprs}
\end{figure}

We construct a set of SRPS by considering a $\nh$-dimensional vector and populating $\nh^D$ entries with independent Gaussian random numbers with zero mean and standard deviation $\nh^{-D/2}$ where $0<D<1$. The construction ensures that the average generalised IPR of the state $\sum_{\ia\ib}\braket{|\psi_{\ia\ib}|^{2q}}\sim \nh^{-D(q-1)}$ and the state therefore has fractal statistics on Hilbert space. In order to compute $\braket{C(r_A,r_B)}$ we can again use Wick's probability theorem to average over the independent Gaussian random numbers and obtain
\eq{
	\braket{C(r_A,r_B)} = &\delta_{r_A0}\sum_{\ia}\sum_{\substack{\ib\jb:\\r_{\ib\jb}=r_B}}\braket{|\psi_{\ia\ib}|^2|\psi_{\ia\jb}|^2}+\nonumber\\ &\delta_{r_B0}\sum_{\ib}\sum_{\substack{\ia\ja:\\r_{\ia\ja}=r_A}}\braket{|\psi_{\ia\ib}|^2|\psi_{\ja\ib}|^2}\,.
	\label{eq:crarbsprs}
}
The above form already implies that $\braket{C(r_A,r_B)}$ is finite only if $r_A
= 0$ or $r_B= 0$ which is in stark contrast to that of MBL states and already indicates why SRPS are not compatible with area-law entanglement. To make this concrete, we next calculate $\braket{C(r_A,r_B)}$ explicitly. Note that the sparse construction of the states implies that several of the summands in Eq.~\ref{eq:crarbsprs} are vanishing. In order to estimate the number of non-zero elements, the following geometrical picture is useful, see Fig.~\ref{fig:sprs}. Consider $\{i_{A(B)}\}$ to be sets each with $\sqrt{\nh}$ elements. Every non-zero $\psi_{\ia\ib}$ corresponds to a link between the respective $\ia$ and $\ib$. By construction there are $\nh^D$ such links and hence $\nh^D$ link terminations in each set.

For $D<1/2$, $\nh^D\ll\sqrt{\nh}$ and hence the link terminations sample the sets sparsely. As such, one can approximate that each element of the sets has only one or zero links connected to it. This implies that any $i_{A(B)}$ is either connected to a unique $i_{B(A)}$ or not connected to anything. In this case, Eq.~\ref{eq:crarbsprs} becomes
\eq{
	\braket{C(r_A,r_B)}= & \delta_{r_A0}\delta_{r_B0}\sideset{}{'}\sum_{\mathclap{\ia,\ib}}\braket{|\psi_{\ia\ib}|^4} =3\delta_{r_A0}\delta_{r_B0}\nh^{-D}\,,
	\label{eq:sprsDl}
}
where the primed summation denotes that sum is over the $\nh^D$ non-zero elements. The result derived above in Eq.~\ref{eq:sprsDl} is indeed corroborated by numerical results shown in Fig.~\ref{fig:sprs}(top right). From Eq.~\ref{eq:sprsDl}, $\braket{\pu} = 3\nh^{-D}$ and hence $S^A_{2,\mathrm{ann}}=DL\ln 2$ which is volume law.

For $D>1/2$, $\nh^D\gg\sqrt{\nh}$ and hence the link terminations sample the sets densely. Each $i_{A(B)}$ has a large number of links connected to it. Therefore each term in the summation over $i_{A(B)}$ in the first (second) line of Eq.~\ref{eq:crarbsprs} contributes. 
At the same time, $\phi_{\ia}\equiv\sum_{\ib}|\psi_{\ia\ib}|^2$ is the $\ia^\mathrm{th}$ diagonal matrix element of $\rho_A$. Since each $\phi_{\ia}$ is a sum of exponentially large in $L$ in random numbers whose distribution's width decays exponentially in $L$, it is reasonable to assume that all $\phi_{\ia}$ are same on average. Normalisation of $\rho_A$ ensures that $\braket{\phi_{\ia}}\sim \nh^{-1/2}$. Moreover, one can also assume equipartition on average for the sum in $\phi_{\ia}$. Using these in Eq.~\ref{eq:crarbsprs}, we find
\eq{
	\braket{C(r_A,r_B)}\approx\frac{\delta_{r_A0}\binom{L/2}{r_B}+\delta_{r_B0}\binom{L/2}{r_A}}{\nh}+\frac{3\delta_{r_A0}\delta_{r_B0}}{\nh^D}\,,
	\label{eq:sprsDg}
}
which is in perfect agreement with the exact numerical result shown in Fig.~\ref{fig:sprs}(bottom right). From Eq.~\ref{eq:sprsDg}, $\braket{\pu}\approx 2\nh^{-1/2}$ and hence $S^A_{2,\mathrm{ann}}\approx L (\ln 2)/2$ which is the same result as for fully extended ergodic states. Within the realm of SRPS, we therefore have $S^A_{2,\mathrm{ann}}=\mathcal{D}L\ln2$ with $\mathcal{D}=D$ for $D\leq1/2$ and $\mathcal{D}=1/2$ for ${1/2<D<1}$, consistent with earlier work~\cite{detomasi2020multifractality}.

The analysis for SRPS shows that while they exhibit (multi)fractal statistics on Hilbert space, exactly like MBL states, the behaviour of $\braket{C(r_A,r_B)}$ is completely different. For MBL states $\braket{C(r_A,r_B)}$ is finite for and sharply peaked at some $r_A,r_B\sim L$. On the contrary, for SRPS (and also for ergodic states) $\braket{C(r_A,r_B)}$ is finite only if $r_A=0$ or $r_B=0$. The calculations above show that this difference arises precisely due to the state amplitudes on Hilbert space being \emph{independent} random numbers for SRPS and ergodic states but for MBL states, the amplitudes are strongly correlated. This can be intuitively understood as if the MBL state for a system of size $L$ is interpreted approximately as a tensor product of $L/\xi$ states corresponding to subsystems of size $\xi$ (thinking of $\xi$ as a proxy for localisation length), then the state is specified using $~e^{\xi}L/\xi$ random numbers which is $\ll \nh^D$. The amplitudes are then naturally correlated.

To summarise, we showed that the purity, a measure of the bipartite entanglement, of a quantum state can be mapped onto a specific four-point correlation function of the amplitudes on Hilbert space, $C(r_A,r_B)$ defined in Eq.~\ref{eq:CrArB}, which naturally goes beyond multifractality. Using numerical calculations for a disordered spin-chain and analytic calculations in limiting cases, we showed that this correlation function is fundamentally different between volume-law entangled ergodic states and area-law entangled MBL states and as an aside, it can be possibly used to diagnose the many-body localisation transition. Finally we calculated $C(r_A,r_B)$ for SRPS which have fractal statistics on the Hilbert space but are devoid of the correlations present in MBL states and exhibit volume-law bipartite entanglement. This allowed us to show that these correlations are at the heart of the coexistence of multifractality on Hilbert space and area-law bipartite entanglement for MBL states, thus answering the fundamental question which motivated this work.

Our analysis in this work centred on eigenstates of an MBL system. The next natural step is consider dynamical versions of these correlations. The motivation for this will be to understand the logarithmic growth of entanglement in the MBL phase~\cite{bardarson2012unbounded,serbyn2013universal} from a Hilbert-space perspective. This might also allow us to theoretically understand the role of resonances~\cite{garratt2021resonances,morningstar2021avalanches,crowley2022constructive,garratt2022resonant} in the dynamics of entanglement in the MBL phase.
While this work focused on area-law entanglement of MBL eigenstates, a question of immediate future interest is the fate of these correlations dynamically across measurement-induced entanglement transitions~\cite{li2018quantum,*li2019measurement,skinner2019measurement}. In such a setting multifractality is known to set in for any finite measurement rate~\cite{sierant2022universal}. The ideas could possibly be extended to measurement-induced transitions in operator entanglement relevant for all-to-all coupled models~\cite{nahum2021measurement}.

\begin{acknowledgments}
The author thanks A.~Lazarides for useful comments on the manuscript. This work was supported by an ICTS-Simons Early Career Faculty Fellowship via a grant from the Simons Foundation (677895, R.G.). All numerical computations were performed on the $\texttt{CONTRA}$ cluster at ICTS-TIFR.
\end{acknowledgments}

\bibliography{refs}

\begin{thebibliography}{48}%
\makeatletter
\providecommand \@ifxundefined [1]{%
 \@ifx{#1\undefined}
}%
\providecommand \@ifnum [1]{%
 \ifnum #1\expandafter \@firstoftwo
 \else \expandafter \@secondoftwo
 \fi
}%
\providecommand \@ifx [1]{%
 \ifx #1\expandafter \@firstoftwo
 \else \expandafter \@secondoftwo
 \fi
}%
\providecommand \natexlab [1]{#1}%
\providecommand \enquote  [1]{``#1''}%
\providecommand \bibnamefont  [1]{#1}%
\providecommand \bibfnamefont [1]{#1}%
\providecommand \citenamefont [1]{#1}%
\providecommand \href@noop [0]{\@secondoftwo}%
\providecommand \href [0]{\begingroup \@sanitize@url \@href}%
\providecommand \@href[1]{\@@startlink{#1}\@@href}%
\providecommand \@@href[1]{\endgroup#1\@@endlink}%
\providecommand \@sanitize@url [0]{\catcode `\\12\catcode `\$12\catcode
  `\&12\catcode `\#12\catcode `\^12\catcode `\_12\catcode `\%12\relax}%
\providecommand \@@startlink[1]{}%
\providecommand \@@endlink[0]{}%
\providecommand \url  [0]{\begingroup\@sanitize@url \@url }%
\providecommand \@url [1]{\endgroup\@href {#1}{\urlprefix }}%
\providecommand \urlprefix  [0]{URL }%
\providecommand \Eprint [0]{\href }%
\providecommand \doibase [0]{https://doi.org/}%
\providecommand \selectlanguage [0]{\@gobble}%
\providecommand \bibinfo  [0]{\@secondoftwo}%
\providecommand \bibfield  [0]{\@secondoftwo}%
\providecommand \translation [1]{[#1]}%
\providecommand \BibitemOpen [0]{}%
\providecommand \bibitemStop [0]{}%
\providecommand \bibitemNoStop [0]{.\EOS\space}%
\providecommand \EOS [0]{\spacefactor3000\relax}%
\providecommand \BibitemShut  [1]{\csname bibitem#1\endcsname}%
\let\auto@bib@innerbib\@empty
\bibitem [{\citenamefont {Gornyi}\ \emph {et~al.}(2005)\citenamefont {Gornyi},
  \citenamefont {Mirlin},\ and\ \citenamefont
  {Polyakov}}]{gornyi2005interacting}%
  \BibitemOpen
  \bibfield  {author} {\bibinfo {author} {\bibfnamefont {I.~V.}\ \bibnamefont
  {Gornyi}}, \bibinfo {author} {\bibfnamefont {A.~D.}\ \bibnamefont {Mirlin}},\
  and\ \bibinfo {author} {\bibfnamefont {D.~G.}\ \bibnamefont {Polyakov}},\
  }\bibfield  {title} {\bibinfo {title} {Interacting electrons in disordered
  wires: Anderson localization and low-${T}$ transport},\ }\href
  {https://doi.org/10.1103/PhysRevLett.95.206603} {\bibfield  {journal}
  {\bibinfo  {journal} {Phys. Rev. Lett.}\ }\textbf {\bibinfo {volume} {95}},\
  \bibinfo {pages} {206603} (\bibinfo {year} {2005})}\BibitemShut {NoStop}%
\bibitem [{\citenamefont {Basko}\ \emph {et~al.}(2006)\citenamefont {Basko},
  \citenamefont {Aleiner},\ and\ \citenamefont {Altshuler}}]{basko2006metal}%
  \BibitemOpen
  \bibfield  {author} {\bibinfo {author} {\bibfnamefont {D.~M.}\ \bibnamefont
  {Basko}}, \bibinfo {author} {\bibfnamefont {I.~L.}\ \bibnamefont {Aleiner}},\
  and\ \bibinfo {author} {\bibfnamefont {B.~L.}\ \bibnamefont {Altshuler}},\
  }\bibfield  {title} {\bibinfo {title} {Metal--insulator transition in a
  weakly interacting many-electron system with localized single-particle
  states},\ }\href
  {http://www.sciencedirect.com/science/article/pii/S0003491605002630}
  {\bibfield  {journal} {\bibinfo  {journal} {Annals of {P}hysics}\ }\textbf
  {\bibinfo {volume} {321}},\ \bibinfo {pages} {1126} (\bibinfo {year}
  {2006})}\BibitemShut {NoStop}%
\bibitem [{\citenamefont {Oganesyan}\ and\ \citenamefont
  {Huse}(2007)}]{oganesyan2007localisation}%
  \BibitemOpen
  \bibfield  {author} {\bibinfo {author} {\bibfnamefont {V.}~\bibnamefont
  {Oganesyan}}\ and\ \bibinfo {author} {\bibfnamefont {D.~A.}\ \bibnamefont
  {Huse}},\ }\bibfield  {title} {\bibinfo {title} {Localization of interacting
  fermions at high temperature},\ }\href
  {https://doi.org/10.1103/PhysRevB.75.155111} {\bibfield  {journal} {\bibinfo
  {journal} {Phys. Rev. B}\ }\textbf {\bibinfo {volume} {75}},\ \bibinfo
  {pages} {155111} (\bibinfo {year} {2007})}\BibitemShut {NoStop}%
\bibitem [{\citenamefont {Pal}\ and\ \citenamefont {Huse}(2010)}]{pal2010many}%
  \BibitemOpen
  \bibfield  {author} {\bibinfo {author} {\bibfnamefont {A.}~\bibnamefont
  {Pal}}\ and\ \bibinfo {author} {\bibfnamefont {D.~A.}\ \bibnamefont {Huse}},\
  }\bibfield  {title} {\bibinfo {title} {Many-body localization phase
  transition},\ }\href {https://doi.org/10.1103/PhysRevB.82.174411} {\bibfield
  {journal} {\bibinfo  {journal} {Phys. Rev. B}\ }\textbf {\bibinfo {volume}
  {82}},\ \bibinfo {pages} {174411} (\bibinfo {year} {2010})}\BibitemShut
  {NoStop}%
\bibitem [{\citenamefont {Imbrie}(2016)}]{imbrie2016many}%
  \BibitemOpen
  \bibfield  {author} {\bibinfo {author} {\bibfnamefont {J.~Z.}\ \bibnamefont
  {Imbrie}},\ }\bibfield  {title} {\bibinfo {title} {On many-body localization
  for quantum spin chains},\ }\href
  {https://link.springer.com/article/10.1007%2Fs10955-016-1508-x} {\bibfield
  {journal} {\bibinfo  {journal} {J. Stat. Phys.}\ }\textbf {\bibinfo {volume}
  {163}},\ \bibinfo {pages} {998} (\bibinfo {year} {2016})}\BibitemShut
  {NoStop}%
\bibitem [{\citenamefont {Nandkishore}\ and\ \citenamefont
  {Huse}(2015)}]{nandkishore2015many}%
  \BibitemOpen
  \bibfield  {author} {\bibinfo {author} {\bibfnamefont {R.}~\bibnamefont
  {Nandkishore}}\ and\ \bibinfo {author} {\bibfnamefont {D.~A.}\ \bibnamefont
  {Huse}},\ }\bibfield  {title} {\bibinfo {title} {Many-body localization and
  thermalization in quantum statistical mechanics},\ }\href
  {https://doi.org/10.1146/annurev-conmatphys-031214-014726} {\bibfield
  {journal} {\bibinfo  {journal} {Annu. Rev. Condens. Matter Phys.}\ }\textbf
  {\bibinfo {volume} {6}},\ \bibinfo {pages} {15} (\bibinfo {year}
  {2015})}\BibitemShut {NoStop}%
\bibitem [{\citenamefont {Abanin}\ and\ \citenamefont
  {Papi{\'c}}(2017)}]{abanin2017recent}%
  \BibitemOpen
  \bibfield  {author} {\bibinfo {author} {\bibfnamefont {D.~A.}\ \bibnamefont
  {Abanin}}\ and\ \bibinfo {author} {\bibfnamefont {Z.}~\bibnamefont
  {Papi{\'c}}},\ }\bibfield  {title} {\bibinfo {title} {Recent progress in
  many-body localization},\ }\href {http://dx.doi.org/10.1002/andp.201700169}
  {\bibfield  {journal} {\bibinfo  {journal} {Annalen der Physik}\ }\textbf
  {\bibinfo {volume} {529}},\ \bibinfo {pages} {1700169} (\bibinfo {year}
  {2017})}\BibitemShut {NoStop}%
\bibitem [{\citenamefont {Alet}\ and\ \citenamefont
  {Laflorencie}(2018)}]{alet2018many}%
  \BibitemOpen
  \bibfield  {author} {\bibinfo {author} {\bibfnamefont {F.}~\bibnamefont
  {Alet}}\ and\ \bibinfo {author} {\bibfnamefont {N.}~\bibnamefont
  {Laflorencie}},\ }\bibfield  {title} {\bibinfo {title} {Many-body
  localization: an introduction and selected topics},\ }\href
  {https://doi.org/https://doi.org/10.1016/j.crhy.2018.03.003} {\bibfield
  {journal} {\bibinfo  {journal} {Comptes Rendus Physique}\ }\textbf {\bibinfo
  {volume} {19}},\ \bibinfo {pages} {498} (\bibinfo {year} {2018})}\BibitemShut
  {NoStop}%
\bibitem [{\citenamefont {Abanin}\ \emph {et~al.}(2019)\citenamefont {Abanin},
  \citenamefont {Altman}, \citenamefont {Bloch},\ and\ \citenamefont
  {Serbyn}}]{abanin2019colloquium}%
  \BibitemOpen
  \bibfield  {author} {\bibinfo {author} {\bibfnamefont {D.~A.}\ \bibnamefont
  {Abanin}}, \bibinfo {author} {\bibfnamefont {E.}~\bibnamefont {Altman}},
  \bibinfo {author} {\bibfnamefont {I.}~\bibnamefont {Bloch}},\ and\ \bibinfo
  {author} {\bibfnamefont {M.}~\bibnamefont {Serbyn}},\ }\bibfield  {title}
  {\bibinfo {title} {Colloquium: Many-body localization, thermalization, and
  entanglement},\ }\href {https://doi.org/10.1103/RevModPhys.91.021001}
  {\bibfield  {journal} {\bibinfo  {journal} {Rev. Mod. Phys.}\ }\textbf
  {\bibinfo {volume} {91}},\ \bibinfo {pages} {021001} (\bibinfo {year}
  {2019})}\BibitemShut {NoStop}%
\bibitem [{\citenamefont {Serbyn}\ \emph
  {et~al.}(2013{\natexlab{a}})\citenamefont {Serbyn}, \citenamefont
  {Papi\ifmmode~\acute{c}\else \'{c}\fi{}},\ and\ \citenamefont
  {Abanin}}]{serbyn2013local}%
  \BibitemOpen
  \bibfield  {author} {\bibinfo {author} {\bibfnamefont {M.}~\bibnamefont
  {Serbyn}}, \bibinfo {author} {\bibfnamefont {Z.}~\bibnamefont
  {Papi\ifmmode~\acute{c}\else \'{c}\fi{}}},\ and\ \bibinfo {author}
  {\bibfnamefont {D.~A.}\ \bibnamefont {Abanin}},\ }\bibfield  {title}
  {\bibinfo {title} {Local conservation laws and the structure of the many-body
  localized states},\ }\href {https://doi.org/10.1103/PhysRevLett.111.127201}
  {\bibfield  {journal} {\bibinfo  {journal} {Phys. Rev. Lett.}\ }\textbf
  {\bibinfo {volume} {111}},\ \bibinfo {pages} {127201} (\bibinfo {year}
  {2013}{\natexlab{a}})}\BibitemShut {NoStop}%
\bibitem [{\citenamefont {Bauer}\ and\ \citenamefont
  {Nayak}(2013)}]{bauer2013area}%
  \BibitemOpen
  \bibfield  {author} {\bibinfo {author} {\bibfnamefont {B.}~\bibnamefont
  {Bauer}}\ and\ \bibinfo {author} {\bibfnamefont {C.}~\bibnamefont {Nayak}},\
  }\bibfield  {title} {\bibinfo {title} {Area laws in a many-body localized
  state and its implications for topological order},\ }\href
  {https://doi.org/10.1088/1742-5468/2013/09/P09005} {\bibfield  {journal}
  {\bibinfo  {journal} {J. Stat. Mech.}\ }\textbf {\bibinfo {volume} {2013}},\
  \bibinfo {pages} {P09005} (\bibinfo {year} {2013})}\BibitemShut {NoStop}%
\bibitem [{\citenamefont {Luitz}\ \emph {et~al.}(2015)\citenamefont {Luitz},
  \citenamefont {Laflorencie},\ and\ \citenamefont {Alet}}]{luitz2015many}%
  \BibitemOpen
  \bibfield  {author} {\bibinfo {author} {\bibfnamefont {D.~J.}\ \bibnamefont
  {Luitz}}, \bibinfo {author} {\bibfnamefont {N.}~\bibnamefont {Laflorencie}},\
  and\ \bibinfo {author} {\bibfnamefont {F.}~\bibnamefont {Alet}},\ }\bibfield
  {title} {\bibinfo {title} {Many-body localization edge in the random-field
  {H}eisenberg chain},\ }\href {https://doi.org/10.1103/PhysRevB.91.081103}
  {\bibfield  {journal} {\bibinfo  {journal} {Phys. Rev. B}\ }\textbf {\bibinfo
  {volume} {91}},\ \bibinfo {pages} {081103} (\bibinfo {year}
  {2015})}\BibitemShut {NoStop}%
\bibitem [{\citenamefont {Geraedts}\ \emph {et~al.}(2016)\citenamefont
  {Geraedts}, \citenamefont {Nandkishore},\ and\ \citenamefont
  {Regnault}}]{geraedts2016many}%
  \BibitemOpen
  \bibfield  {author} {\bibinfo {author} {\bibfnamefont {S.~D.}\ \bibnamefont
  {Geraedts}}, \bibinfo {author} {\bibfnamefont {R.}~\bibnamefont
  {Nandkishore}},\ and\ \bibinfo {author} {\bibfnamefont {N.}~\bibnamefont
  {Regnault}},\ }\bibfield  {title} {\bibinfo {title} {Many-body localization
  and thermalization: Insights from the entanglement spectrum},\ }\href
  {https://doi.org/10.1103/PhysRevB.93.174202} {\bibfield  {journal} {\bibinfo
  {journal} {Phys. Rev. B}\ }\textbf {\bibinfo {volume} {93}},\ \bibinfo
  {pages} {174202} (\bibinfo {year} {2016})}\BibitemShut {NoStop}%
\bibitem [{\citenamefont {Geraedts}\ \emph {et~al.}(2017)\citenamefont
  {Geraedts}, \citenamefont {Regnault},\ and\ \citenamefont
  {Nandkishore}}]{geraedts2017characterising}%
  \BibitemOpen
  \bibfield  {author} {\bibinfo {author} {\bibfnamefont {S.~D.}\ \bibnamefont
  {Geraedts}}, \bibinfo {author} {\bibfnamefont {N.}~\bibnamefont {Regnault}},\
  and\ \bibinfo {author} {\bibfnamefont {R.~M.}\ \bibnamefont {Nandkishore}},\
  }\bibfield  {title} {\bibinfo {title} {Characterizing the many-body
  localization transition using the entanglement spectrum},\ }\href
  {https://doi.org/10.1088/1367-2630/aa93a5} {\bibfield  {journal} {\bibinfo
  {journal} {New J. Phys.}\ }\textbf {\bibinfo {volume} {19}},\ \bibinfo
  {pages} {113021} (\bibinfo {year} {2017})}\BibitemShut {NoStop}%
\bibitem [{\citenamefont {Khemani}\ \emph {et~al.}(2017)\citenamefont
  {Khemani}, \citenamefont {Lim}, \citenamefont {Sheng},\ and\ \citenamefont
  {Huse}}]{khemani2017critical}%
  \BibitemOpen
  \bibfield  {author} {\bibinfo {author} {\bibfnamefont {V.}~\bibnamefont
  {Khemani}}, \bibinfo {author} {\bibfnamefont {S.~P.}\ \bibnamefont {Lim}},
  \bibinfo {author} {\bibfnamefont {D.~N.}\ \bibnamefont {Sheng}},\ and\
  \bibinfo {author} {\bibfnamefont {D.~A.}\ \bibnamefont {Huse}},\ }\bibfield
  {title} {\bibinfo {title} {Critical properties of the many-body localization
  transition},\ }\href {https://doi.org/10.1103/PhysRevX.7.021013} {\bibfield
  {journal} {\bibinfo  {journal} {Phys. Rev. X}\ }\textbf {\bibinfo {volume}
  {7}},\ \bibinfo {pages} {021013} (\bibinfo {year} {2017})}\BibitemShut
  {NoStop}%
\bibitem [{Note1()}]{Note1}%
  \BibitemOpen
  \bibinfo {note} {Of course in the extreme limit of infinite disorder
  strength, the eigenstates are product states and the fractal dimension is
  zero.}\BibitemShut {Stop}%
\bibitem [{\citenamefont {De~Luca}\ and\ \citenamefont
  {Scardicchio}(2013)}]{deluca2013ergodicity}%
  \BibitemOpen
  \bibfield  {author} {\bibinfo {author} {\bibfnamefont {A.}~\bibnamefont
  {De~Luca}}\ and\ \bibinfo {author} {\bibfnamefont {A.}~\bibnamefont
  {Scardicchio}},\ }\bibfield  {title} {\bibinfo {title} {Ergodicity breaking
  in a model showing many-body localization},\ }\href
  {https://doi.org/10.1209/0295-5075/101/37003} {\bibfield  {journal} {\bibinfo
   {journal} {Europhys. Lett.}\ }\textbf {\bibinfo {volume} {101}},\ \bibinfo
  {pages} {37003} (\bibinfo {year} {2013})}\BibitemShut {NoStop}%
\bibitem [{\citenamefont {Mac\'e}\ \emph {et~al.}(2019)\citenamefont {Mac\'e},
  \citenamefont {Alet},\ and\ \citenamefont
  {Laflorencie}}]{mace2019multifractal}%
  \BibitemOpen
  \bibfield  {author} {\bibinfo {author} {\bibfnamefont {N.}~\bibnamefont
  {Mac\'e}}, \bibinfo {author} {\bibfnamefont {F.}~\bibnamefont {Alet}},\ and\
  \bibinfo {author} {\bibfnamefont {N.}~\bibnamefont {Laflorencie}},\
  }\bibfield  {title} {\bibinfo {title} {Multifractal scalings across the
  many-body localization transition},\ }\href
  {https://doi.org/10.1103/PhysRevLett.123.180601} {\bibfield  {journal}
  {\bibinfo  {journal} {Phys. Rev. Lett.}\ }\textbf {\bibinfo {volume} {123}},\
  \bibinfo {pages} {180601} (\bibinfo {year} {2019})}\BibitemShut {NoStop}%
\bibitem [{\citenamefont {De~Tomasi}\ \emph {et~al.}(2021)\citenamefont
  {De~Tomasi}, \citenamefont {Khaymovich}, \citenamefont {Pollmann},\ and\
  \citenamefont {Warzel}}]{detomasi2020rare}%
  \BibitemOpen
  \bibfield  {author} {\bibinfo {author} {\bibfnamefont {G.}~\bibnamefont
  {De~Tomasi}}, \bibinfo {author} {\bibfnamefont {I.~M.}\ \bibnamefont
  {Khaymovich}}, \bibinfo {author} {\bibfnamefont {F.}~\bibnamefont
  {Pollmann}},\ and\ \bibinfo {author} {\bibfnamefont {S.}~\bibnamefont
  {Warzel}},\ }\bibfield  {title} {\bibinfo {title} {Rare thermal bubbles at
  the many-body localization transition from the fock space point of view},\
  }\href {https://doi.org/10.1103/PhysRevB.104.024202} {\bibfield  {journal}
  {\bibinfo  {journal} {Phys. Rev. B}\ }\textbf {\bibinfo {volume} {104}},\
  \bibinfo {pages} {024202} (\bibinfo {year} {2021})}\BibitemShut {NoStop}%
\bibitem [{\citenamefont {Roy}\ and\ \citenamefont
  {Logan}(2021)}]{roy2021fockspace}%
  \BibitemOpen
  \bibfield  {author} {\bibinfo {author} {\bibfnamefont {S.}~\bibnamefont
  {Roy}}\ and\ \bibinfo {author} {\bibfnamefont {D.~E.}\ \bibnamefont
  {Logan}},\ }\bibfield  {title} {\bibinfo {title} {Fock-space anatomy of
  eigenstates across the many-body localization transition},\ }\href
  {https://doi.org/10.1103/PhysRevB.104.174201} {\bibfield  {journal} {\bibinfo
   {journal} {Phys. Rev. B}\ }\textbf {\bibinfo {volume} {104}},\ \bibinfo
  {pages} {174201} (\bibinfo {year} {2021})}\BibitemShut {NoStop}%
\bibitem [{\citenamefont {De~Tomasi}\ and\ \citenamefont
  {Khaymovich}(2020)}]{detomasi2020multifractality}%
  \BibitemOpen
  \bibfield  {author} {\bibinfo {author} {\bibfnamefont {G.}~\bibnamefont
  {De~Tomasi}}\ and\ \bibinfo {author} {\bibfnamefont {I.~M.}\ \bibnamefont
  {Khaymovich}},\ }\bibfield  {title} {\bibinfo {title} {Multifractality meets
  entanglement: Relation for nonergodic extended states},\ }\href
  {https://doi.org/10.1103/PhysRevLett.124.200602} {\bibfield  {journal}
  {\bibinfo  {journal} {Phys. Rev. Lett.}\ }\textbf {\bibinfo {volume} {124}},\
  \bibinfo {pages} {200602} (\bibinfo {year} {2020})}\BibitemShut {NoStop}%
\bibitem [{\citenamefont {Lubkin}(1978)}]{lubkin1978entropy}%
  \BibitemOpen
  \bibfield  {author} {\bibinfo {author} {\bibfnamefont {E.}~\bibnamefont
  {Lubkin}},\ }\bibfield  {title} {\bibinfo {title} {Entropy of an $n$‐system
  from its correlation with a $k$‐reservoir},\ }\href
  {https://doi.org/10.1063/1.523763} {\bibfield  {journal} {\bibinfo  {journal}
  {J. Math. Phys.}\ }\textbf {\bibinfo {volume} {19}},\ \bibinfo {pages} {1028}
  (\bibinfo {year} {1978})}\BibitemShut {NoStop}%
\bibitem [{\citenamefont {Page}(1993)}]{page1993average}%
  \BibitemOpen
  \bibfield  {author} {\bibinfo {author} {\bibfnamefont {D.~N.}\ \bibnamefont
  {Page}},\ }\bibfield  {title} {\bibinfo {title} {Average entropy of a
  subsystem},\ }\href {https://doi.org/10.1103/PhysRevLett.71.1291} {\bibfield
  {journal} {\bibinfo  {journal} {Phys. Rev. Lett.}\ }\textbf {\bibinfo
  {volume} {71}},\ \bibinfo {pages} {1291} (\bibinfo {year}
  {1993})}\BibitemShut {NoStop}%
\bibitem [{\citenamefont {Lu}\ and\ \citenamefont
  {Grover}(2019)}]{lu2019renyi}%
  \BibitemOpen
  \bibfield  {author} {\bibinfo {author} {\bibfnamefont {T.-C.}\ \bibnamefont
  {Lu}}\ and\ \bibinfo {author} {\bibfnamefont {T.}~\bibnamefont {Grover}},\
  }\bibfield  {title} {\bibinfo {title} {Renyi entropy of chaotic
  eigenstates},\ }\href {https://doi.org/10.1103/PhysRevE.99.032111} {\bibfield
   {journal} {\bibinfo  {journal} {Phys. Rev. E}\ }\textbf {\bibinfo {volume}
  {99}},\ \bibinfo {pages} {032111} (\bibinfo {year} {2019})}\BibitemShut
  {NoStop}%
\bibitem [{\citenamefont {Monthus}(2016)}]{monthus2016manybody}%
  \BibitemOpen
  \bibfield  {author} {\bibinfo {author} {\bibfnamefont {C.}~\bibnamefont
  {Monthus}},\ }\bibfield  {title} {\bibinfo {title} {Many-body-localization
  transition in the strong disorder limit: Entanglement entropy from the
  statistics of rare extensive resonances},\ }\href
  {https://doi.org/10.3390/e18040122} {\bibfield  {journal} {\bibinfo
  {journal} {Entropy}\ }\textbf {\bibinfo {volume} {18}},\ \bibinfo {pages}
  {122} (\bibinfo {year} {2016})}\BibitemShut {NoStop}%
\bibitem [{\citenamefont {Abanin}\ \emph {et~al.}(2021)\citenamefont {Abanin},
  \citenamefont {Bardarson}, \citenamefont {{De Tomasi}}, \citenamefont
  {Gopalakrishnan}, \citenamefont {Khemani}, \citenamefont {Parameswaran},
  \citenamefont {Pollmann}, \citenamefont {Potter}, \citenamefont {Serbyn},\
  and\ \citenamefont {Vasseur}}]{abanin2021distinguishing}%
  \BibitemOpen
  \bibfield  {author} {\bibinfo {author} {\bibfnamefont {D.}~\bibnamefont
  {Abanin}}, \bibinfo {author} {\bibfnamefont {J.}~\bibnamefont {Bardarson}},
  \bibinfo {author} {\bibfnamefont {G.}~\bibnamefont {{De Tomasi}}}, \bibinfo
  {author} {\bibfnamefont {S.}~\bibnamefont {Gopalakrishnan}}, \bibinfo
  {author} {\bibfnamefont {V.}~\bibnamefont {Khemani}}, \bibinfo {author}
  {\bibfnamefont {S.}~\bibnamefont {Parameswaran}}, \bibinfo {author}
  {\bibfnamefont {F.}~\bibnamefont {Pollmann}}, \bibinfo {author}
  {\bibfnamefont {A.}~\bibnamefont {Potter}}, \bibinfo {author} {\bibfnamefont
  {M.}~\bibnamefont {Serbyn}},\ and\ \bibinfo {author} {\bibfnamefont
  {R.}~\bibnamefont {Vasseur}},\ }\bibfield  {title} {\bibinfo {title}
  {Distinguishing localization from chaos: Challenges in finite-size systems},\
  }\href {https://doi.org/https://doi.org/10.1016/j.aop.2021.168415} {\bibfield
   {journal} {\bibinfo  {journal} {Ann. Phys.}\ }\textbf {\bibinfo {volume}
  {427}},\ \bibinfo {pages} {168415} (\bibinfo {year} {2021})}\BibitemShut
  {NoStop}%
\bibitem [{\citenamefont {\ifmmode~\check{S}\else \v{S}\fi{}untajs}\ \emph
  {et~al.}(2020)\citenamefont {\ifmmode~\check{S}\else \v{S}\fi{}untajs},
  \citenamefont {Bon\ifmmode~\check{c}\else \v{c}\fi{}a}, \citenamefont
  {Prosen},\ and\ \citenamefont {Vidmar}}]{suntajs2020quantum}%
  \BibitemOpen
  \bibfield  {author} {\bibinfo {author} {\bibfnamefont {J.}~\bibnamefont
  {\ifmmode~\check{S}\else \v{S}\fi{}untajs}}, \bibinfo {author} {\bibfnamefont
  {J.}~\bibnamefont {Bon\ifmmode~\check{c}\else \v{c}\fi{}a}}, \bibinfo
  {author} {\bibfnamefont {T.}~\bibnamefont {Prosen}},\ and\ \bibinfo {author}
  {\bibfnamefont {L.}~\bibnamefont {Vidmar}},\ }\bibfield  {title} {\bibinfo
  {title} {Quantum chaos challenges many-body localization},\ }\href
  {https://doi.org/10.1103/PhysRevE.102.062144} {\bibfield  {journal} {\bibinfo
   {journal} {Phys. Rev. E}\ }\textbf {\bibinfo {volume} {102}},\ \bibinfo
  {pages} {062144} (\bibinfo {year} {2020})}\BibitemShut {NoStop}%
\bibitem [{\citenamefont {Morningstar}\ \emph {et~al.}(2022)\citenamefont
  {Morningstar}, \citenamefont {Colmenarez}, \citenamefont {Khemani},
  \citenamefont {Luitz},\ and\ \citenamefont
  {Huse}}]{morningstar2021avalanches}%
  \BibitemOpen
  \bibfield  {author} {\bibinfo {author} {\bibfnamefont {A.}~\bibnamefont
  {Morningstar}}, \bibinfo {author} {\bibfnamefont {L.}~\bibnamefont
  {Colmenarez}}, \bibinfo {author} {\bibfnamefont {V.}~\bibnamefont {Khemani}},
  \bibinfo {author} {\bibfnamefont {D.~J.}\ \bibnamefont {Luitz}},\ and\
  \bibinfo {author} {\bibfnamefont {D.~A.}\ \bibnamefont {Huse}},\ }\bibfield
  {title} {\bibinfo {title} {Avalanches and many-body resonances in many-body
  localized systems},\ }\href {https://doi.org/10.1103/PhysRevB.105.174205}
  {\bibfield  {journal} {\bibinfo  {journal} {Phys. Rev. B}\ }\textbf {\bibinfo
  {volume} {105}},\ \bibinfo {pages} {174205} (\bibinfo {year}
  {2022})}\BibitemShut {NoStop}%
\bibitem [{\citenamefont {Sels}(2022)}]{sels2022bath}%
  \BibitemOpen
  \bibfield  {author} {\bibinfo {author} {\bibfnamefont {D.}~\bibnamefont
  {Sels}},\ }\bibfield  {title} {\bibinfo {title} {Bath-induced delocalization
  in interacting disordered spin chains},\ }\href
  {https://doi.org/10.1103/PhysRevB.106.L020202} {\bibfield  {journal}
  {\bibinfo  {journal} {Phys. Rev. B}\ }\textbf {\bibinfo {volume} {106}},\
  \bibinfo {pages} {L020202} (\bibinfo {year} {2022})}\BibitemShut {NoStop}%
\bibitem [{\citenamefont {Gopalakrishnan}\ \emph {et~al.}(2015)\citenamefont
  {Gopalakrishnan}, \citenamefont {M\"uller}, \citenamefont {Khemani},
  \citenamefont {Knap}, \citenamefont {Demler},\ and\ \citenamefont
  {Huse}}]{gopalakrishnan2015lowfrequency}%
  \BibitemOpen
  \bibfield  {author} {\bibinfo {author} {\bibfnamefont {S.}~\bibnamefont
  {Gopalakrishnan}}, \bibinfo {author} {\bibfnamefont {M.}~\bibnamefont
  {M\"uller}}, \bibinfo {author} {\bibfnamefont {V.}~\bibnamefont {Khemani}},
  \bibinfo {author} {\bibfnamefont {M.}~\bibnamefont {Knap}}, \bibinfo {author}
  {\bibfnamefont {E.}~\bibnamefont {Demler}},\ and\ \bibinfo {author}
  {\bibfnamefont {D.~A.}\ \bibnamefont {Huse}},\ }\bibfield  {title} {\bibinfo
  {title} {Low-frequency conductivity in many-body localized systems},\ }\href
  {https://doi.org/10.1103/PhysRevB.92.104202} {\bibfield  {journal} {\bibinfo
  {journal} {Phys. Rev. B}\ }\textbf {\bibinfo {volume} {92}},\ \bibinfo
  {pages} {104202} (\bibinfo {year} {2015})}\BibitemShut {NoStop}%
\bibitem [{\citenamefont {Garratt}\ \emph {et~al.}(2021)\citenamefont
  {Garratt}, \citenamefont {Roy},\ and\ \citenamefont
  {Chalker}}]{garratt2021resonances}%
  \BibitemOpen
  \bibfield  {author} {\bibinfo {author} {\bibfnamefont {S.~J.}\ \bibnamefont
  {Garratt}}, \bibinfo {author} {\bibfnamefont {S.}~\bibnamefont {Roy}},\ and\
  \bibinfo {author} {\bibfnamefont {J.~T.}\ \bibnamefont {Chalker}},\
  }\bibfield  {title} {\bibinfo {title} {Local resonances and parametric level
  dynamics in the many-body localized phase},\ }\href
  {https://doi.org/10.1103/PhysRevB.104.184203} {\bibfield  {journal} {\bibinfo
   {journal} {Phys. Rev. B}\ }\textbf {\bibinfo {volume} {104}},\ \bibinfo
  {pages} {184203} (\bibinfo {year} {2021})}\BibitemShut {NoStop}%
\bibitem [{\citenamefont {Crowley}\ and\ \citenamefont
  {Chandran}(2022)}]{crowley2022constructive}%
  \BibitemOpen
  \bibfield  {author} {\bibinfo {author} {\bibfnamefont {P.~J.~D.}\
  \bibnamefont {Crowley}}\ and\ \bibinfo {author} {\bibfnamefont
  {A.}~\bibnamefont {Chandran}},\ }\bibfield  {title} {\bibinfo {title} {{A
  constructive theory of the numerically accessible many-body localized to
  thermal crossover}},\ }\href {https://doi.org/10.21468/SciPostPhys.12.6.201}
  {\bibfield  {journal} {\bibinfo  {journal} {SciPost Phys.}\ }\textbf
  {\bibinfo {volume} {12}},\ \bibinfo {pages} {201} (\bibinfo {year}
  {2022})}\BibitemShut {NoStop}%
\bibitem [{\citenamefont {Garratt}\ and\ \citenamefont
  {Roy}(2022)}]{garratt2022resonant}%
  \BibitemOpen
  \bibfield  {author} {\bibinfo {author} {\bibfnamefont {S.~J.}\ \bibnamefont
  {Garratt}}\ and\ \bibinfo {author} {\bibfnamefont {S.}~\bibnamefont {Roy}},\
  }\href {https://doi.org/10.48550/ARXIV.2202.10482} {\bibinfo {title}
  {Resonant energy scales and local observables in the many-body localised
  phase}} (\bibinfo {year} {2022})\BibitemShut {NoStop}%
\bibitem [{\citenamefont {Long}\ \emph {et~al.}(2022)\citenamefont {Long},
  \citenamefont {Crowley}, \citenamefont {Khemani},\ and\ \citenamefont
  {Chandran}}]{long2022phenomenology}%
  \BibitemOpen
  \bibfield  {author} {\bibinfo {author} {\bibfnamefont {D.~M.}\ \bibnamefont
  {Long}}, \bibinfo {author} {\bibfnamefont {P.~J.~D.}\ \bibnamefont
  {Crowley}}, \bibinfo {author} {\bibfnamefont {V.}~\bibnamefont {Khemani}},\
  and\ \bibinfo {author} {\bibfnamefont {A.}~\bibnamefont {Chandran}},\ }\href
  {https://doi.org/10.48550/ARXIV.2207.05761} {\bibinfo {title} {Phenomenology
  of the prethermal many-body localized regime}} (\bibinfo {year}
  {2022})\BibitemShut {NoStop}%
\bibitem [{\citenamefont {Vidmar}\ \emph {et~al.}(2021)\citenamefont {Vidmar},
  \citenamefont {Krajewski}, \citenamefont {Bon\ifmmode~\check{c}\else
  \v{c}\fi{}a},\ and\ \citenamefont {Mierzejewski}}]{vidmar2021phenomenology}%
  \BibitemOpen
  \bibfield  {author} {\bibinfo {author} {\bibfnamefont {L.}~\bibnamefont
  {Vidmar}}, \bibinfo {author} {\bibfnamefont {B.}~\bibnamefont {Krajewski}},
  \bibinfo {author} {\bibfnamefont {J.}~\bibnamefont
  {Bon\ifmmode~\check{c}\else \v{c}\fi{}a}},\ and\ \bibinfo {author}
  {\bibfnamefont {M.}~\bibnamefont {Mierzejewski}},\ }\bibfield  {title}
  {\bibinfo {title} {Phenomenology of spectral functions in disordered spin
  chains at infinite temperature},\ }\href
  {https://doi.org/10.1103/PhysRevLett.127.230603} {\bibfield  {journal}
  {\bibinfo  {journal} {Phys. Rev. Lett.}\ }\textbf {\bibinfo {volume} {127}},\
  \bibinfo {pages} {230603} (\bibinfo {year} {2021})}\BibitemShut {NoStop}%
\bibitem [{Note2()}]{Note2}%
  \BibitemOpen
  \bibinfo {note} {See supplementary material \cite {supp} for an alternative
  renormalisation}\BibitemShut {NoStop}%
\bibitem [{\citenamefont {Mehta}(2014)}]{mehta2014random}%
  \BibitemOpen
  \bibfield  {author} {\bibinfo {author} {\bibfnamefont {M.~L.}\ \bibnamefont
  {Mehta}},\ }\href {https://books.google.co.in/books?id=\_MjSBQAAQBAJ} {\emph
  {\bibinfo {title} {Random Matrices: Revised and Enlarged Second Edition}}}\
  (\bibinfo  {publisher} {Elsevier Science},\ \bibinfo {year}
  {2014})\BibitemShut {NoStop}%
\bibitem [{\citenamefont {Haake}(2010)}]{haake2010quantum}%
  \BibitemOpen
  \bibfield  {author} {\bibinfo {author} {\bibfnamefont {F.}~\bibnamefont
  {Haake}},\ }\href {https://link.springer.com/book/10.1007/978-3-642-05428-0}
  {\emph {\bibinfo {title} {Quantum Signatures of Chaos}}},\ Springer Series in
  Synergetics\ (\bibinfo  {publisher} {Springer Berlin Heidelberg},\ \bibinfo
  {year} {2010})\BibitemShut {NoStop}%
\bibitem [{\citenamefont {Huse}\ \emph {et~al.}(2014)\citenamefont {Huse},
  \citenamefont {Nandkishore},\ and\ \citenamefont
  {Oganesyan}}]{huse2014phenomenology}%
  \BibitemOpen
  \bibfield  {author} {\bibinfo {author} {\bibfnamefont {D.~A.}\ \bibnamefont
  {Huse}}, \bibinfo {author} {\bibfnamefont {R.}~\bibnamefont {Nandkishore}},\
  and\ \bibinfo {author} {\bibfnamefont {V.}~\bibnamefont {Oganesyan}},\
  }\bibfield  {title} {\bibinfo {title} {Phenomenology of fully
  many-body-localized systems},\ }\href
  {https://doi.org/10.1103/PhysRevB.90.174202} {\bibfield  {journal} {\bibinfo
  {journal} {Phys. Rev. B}\ }\textbf {\bibinfo {volume} {90}},\ \bibinfo
  {pages} {174202} (\bibinfo {year} {2014})}\BibitemShut {NoStop}%
\bibitem [{\citenamefont {Imbrie}\ \emph {et~al.}(2017)\citenamefont {Imbrie},
  \citenamefont {Ros},\ and\ \citenamefont {Scardicchio}}]{imbrie2017local}%
  \BibitemOpen
  \bibfield  {author} {\bibinfo {author} {\bibfnamefont {J.~Z.}\ \bibnamefont
  {Imbrie}}, \bibinfo {author} {\bibfnamefont {V.}~\bibnamefont {Ros}},\ and\
  \bibinfo {author} {\bibfnamefont {A.}~\bibnamefont {Scardicchio}},\
  }\bibfield  {title} {\bibinfo {title} {Local integrals of motion in many-body
  localized systems},\ }\href
  {https://onlinelibrary.wiley.com/doi/full/10.1002/andp.201600278} {\bibfield
  {journal} {\bibinfo  {journal} {Annalen der Physik}\ }\textbf {\bibinfo
  {volume} {529}},\ \bibinfo {pages} {1600278} (\bibinfo {year}
  {2017})}\BibitemShut {NoStop}%
\bibitem [{sup()}]{supp}%
  \BibitemOpen
  \href@noop {} {}\bibinfo {note} {See supplementary material at
  [URL].}\BibitemShut {Stop}%
\bibitem [{\citenamefont {Bardarson}\ \emph {et~al.}(2012)\citenamefont
  {Bardarson}, \citenamefont {Pollmann},\ and\ \citenamefont
  {Moore}}]{bardarson2012unbounded}%
  \BibitemOpen
  \bibfield  {author} {\bibinfo {author} {\bibfnamefont {J.~H.}\ \bibnamefont
  {Bardarson}}, \bibinfo {author} {\bibfnamefont {F.}~\bibnamefont
  {Pollmann}},\ and\ \bibinfo {author} {\bibfnamefont {J.~E.}\ \bibnamefont
  {Moore}},\ }\bibfield  {title} {\bibinfo {title} {Unbounded growth of
  entanglement in models of many-body localization},\ }\href
  {https://doi.org/10.1103/PhysRevLett.109.017202} {\bibfield  {journal}
  {\bibinfo  {journal} {Phys. Rev. Lett.}\ }\textbf {\bibinfo {volume} {109}},\
  \bibinfo {pages} {017202} (\bibinfo {year} {2012})}\BibitemShut {NoStop}%
\bibitem [{\citenamefont {Serbyn}\ \emph
  {et~al.}(2013{\natexlab{b}})\citenamefont {Serbyn}, \citenamefont
  {Papi\ifmmode~\acute{c}\else \'{c}\fi{}},\ and\ \citenamefont
  {Abanin}}]{serbyn2013universal}%
  \BibitemOpen
  \bibfield  {author} {\bibinfo {author} {\bibfnamefont {M.}~\bibnamefont
  {Serbyn}}, \bibinfo {author} {\bibfnamefont {Z.}~\bibnamefont
  {Papi\ifmmode~\acute{c}\else \'{c}\fi{}}},\ and\ \bibinfo {author}
  {\bibfnamefont {D.~A.}\ \bibnamefont {Abanin}},\ }\bibfield  {title}
  {\bibinfo {title} {Universal slow growth of entanglement in interacting
  strongly disordered systems},\ }\href
  {https://doi.org/10.1103/PhysRevLett.110.260601} {\bibfield  {journal}
  {\bibinfo  {journal} {Phys. Rev. Lett.}\ }\textbf {\bibinfo {volume} {110}},\
  \bibinfo {pages} {260601} (\bibinfo {year} {2013}{\natexlab{b}})}\BibitemShut
  {NoStop}%
\bibitem [{\citenamefont {Li}\ \emph {et~al.}(2018)\citenamefont {Li},
  \citenamefont {Chen},\ and\ \citenamefont {Fisher}}]{li2018quantum}%
  \BibitemOpen
  \bibfield  {author} {\bibinfo {author} {\bibfnamefont {Y.}~\bibnamefont
  {Li}}, \bibinfo {author} {\bibfnamefont {X.}~\bibnamefont {Chen}},\ and\
  \bibinfo {author} {\bibfnamefont {M.~P.~A.}\ \bibnamefont {Fisher}},\
  }\bibfield  {title} {\bibinfo {title} {Quantum zeno effect and the many-body
  entanglement transition},\ }\href
  {https://doi.org/10.1103/PhysRevB.98.205136} {\bibfield  {journal} {\bibinfo
  {journal} {Phys. Rev. B}\ }\textbf {\bibinfo {volume} {98}},\ \bibinfo
  {pages} {205136} (\bibinfo {year} {2018})}\BibitemShut {NoStop}%
\bibitem [{\citenamefont {Li}\ \emph {et~al.}(2019)\citenamefont {Li},
  \citenamefont {Chen},\ and\ \citenamefont {Fisher}}]{li2019measurement}%
  \BibitemOpen
  \bibfield  {author} {\bibinfo {author} {\bibfnamefont {Y.}~\bibnamefont
  {Li}}, \bibinfo {author} {\bibfnamefont {X.}~\bibnamefont {Chen}},\ and\
  \bibinfo {author} {\bibfnamefont {M.~P.~A.}\ \bibnamefont {Fisher}},\
  }\bibfield  {title} {\bibinfo {title} {Measurement-driven entanglement
  transition in hybrid quantum circuits},\ }\href
  {https://doi.org/10.1103/PhysRevB.100.134306} {\bibfield  {journal} {\bibinfo
   {journal} {Phys. Rev. B}\ }\textbf {\bibinfo {volume} {100}},\ \bibinfo
  {pages} {134306} (\bibinfo {year} {2019})}\BibitemShut {NoStop}%
\bibitem [{\citenamefont {Skinner}\ \emph {et~al.}(2019)\citenamefont
  {Skinner}, \citenamefont {Ruhman},\ and\ \citenamefont
  {Nahum}}]{skinner2019measurement}%
  \BibitemOpen
  \bibfield  {author} {\bibinfo {author} {\bibfnamefont {B.}~\bibnamefont
  {Skinner}}, \bibinfo {author} {\bibfnamefont {J.}~\bibnamefont {Ruhman}},\
  and\ \bibinfo {author} {\bibfnamefont {A.}~\bibnamefont {Nahum}},\ }\bibfield
   {title} {\bibinfo {title} {Measurement-induced phase transitions in the
  dynamics of entanglement},\ }\href
  {https://doi.org/10.1103/PhysRevX.9.031009} {\bibfield  {journal} {\bibinfo
  {journal} {Phys. Rev. X}\ }\textbf {\bibinfo {volume} {9}},\ \bibinfo {pages}
  {031009} (\bibinfo {year} {2019})}\BibitemShut {NoStop}%
\bibitem [{\citenamefont {Sierant}\ and\ \citenamefont
  {Turkeshi}(2022)}]{sierant2022universal}%
  \BibitemOpen
  \bibfield  {author} {\bibinfo {author} {\bibfnamefont {P.}~\bibnamefont
  {Sierant}}\ and\ \bibinfo {author} {\bibfnamefont {X.}~\bibnamefont
  {Turkeshi}},\ }\bibfield  {title} {\bibinfo {title} {Universal behavior
  beyond multifractality of wave functions at measurement-induced phase
  transitions},\ }\href {https://doi.org/10.1103/PhysRevLett.128.130605}
  {\bibfield  {journal} {\bibinfo  {journal} {Phys. Rev. Lett.}\ }\textbf
  {\bibinfo {volume} {128}},\ \bibinfo {pages} {130605} (\bibinfo {year}
  {2022})}\BibitemShut {NoStop}%
\bibitem [{\citenamefont {Nahum}\ \emph {et~al.}(2021)\citenamefont {Nahum},
  \citenamefont {Roy}, \citenamefont {Skinner},\ and\ \citenamefont
  {Ruhman}}]{nahum2021measurement}%
  \BibitemOpen
  \bibfield  {author} {\bibinfo {author} {\bibfnamefont {A.}~\bibnamefont
  {Nahum}}, \bibinfo {author} {\bibfnamefont {S.}~\bibnamefont {Roy}}, \bibinfo
  {author} {\bibfnamefont {B.}~\bibnamefont {Skinner}},\ and\ \bibinfo {author}
  {\bibfnamefont {J.}~\bibnamefont {Ruhman}},\ }\bibfield  {title} {\bibinfo
  {title} {Measurement and entanglement phase transitions in all-to-all quantum
  circuits, on quantum trees, and in {L}andau-{G}insburg theory},\ }\href
  {https://doi.org/10.1103/PRXQuantum.2.010352} {\bibfield  {journal} {\bibinfo
   {journal} {PRX Quantum}\ }\textbf {\bibinfo {volume} {2}},\ \bibinfo {pages}
  {010352} (\bibinfo {year} {2021})}\BibitemShut {NoStop}%
\end{thebibliography}%

\clearpage

\onecolumngrid

\setcounter{equation}{0}
\setcounter{figure}{0}
\setcounter{page}{1}
\def\theequation{S\arabic{equation}}
\def\thefigure{S\arabic{figure}}
\def\thepage{S\arabic{page}}

\begin{center}
\textbf{{Supplementary Material: Hilbert-space correlations beyond multifractality and bipartite entanglement in many-body localised systems}}
\medskip

Sthitadhi Roy\\
\textit{International Centre for Theoretical Sciences, Tata Institute of Fundamental Research, Bengaluru 560089, India}
\bigskip

\end{center}

\section*{An alternative renormalisation}
\begin{figure}[!h]
\includegraphics[width=0.35\linewidth]{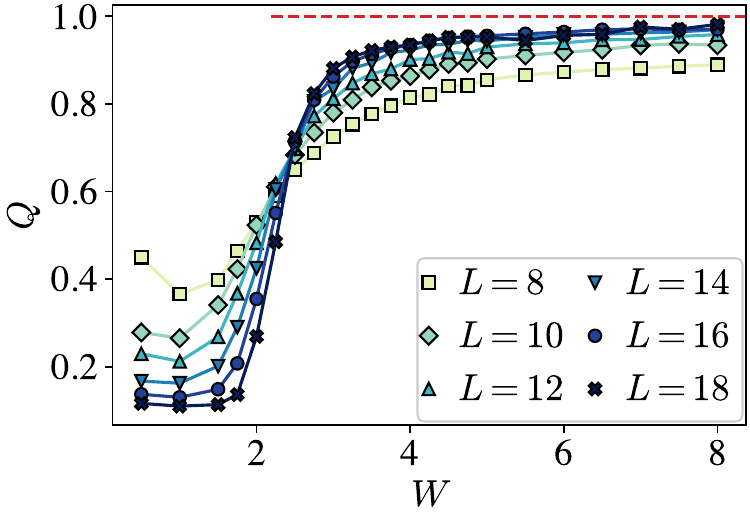}
\caption{The ratio $Q$ defined in Eq.~\ref{eq:Q} as a function of $W$ for several $L$. The red dashed line denotes $Q=1$, the result expected for $W\to\infty$.}
\label{fig:Q}
\end{figure}

In the main text, we showed that in the MBL phase, $\braket{r_Ar_B}$ decays monotonically with $W$. This is because the peak in $\braket{C(r_A,r_B)}$ shifts towards the origin, as the state gets more and more localised in the Hilbert space. In order to account for this we renormalised $\braket{r_Ar_B}$ with $\braket{r_A+r_B}^2$ to define the ratio $R$ in Eq.~\ref{eq:R}. However, one could also define an alternative renormalisation
\eq{
	Q = \frac{\braket{r_A r_B}}{\braket{r_A}\braket{r_B}}\,.
	\label{eq:Q}
}
In the ergodic phase, we again expect $Q\to 0$ as the numerator in Eq.~\ref{eq:Q} itself goes to zero. On the other hand, in the MBL phase since $\braket{r_Ar_B}\sim L^2$ and each of $\braket{r_A}$ and $\braket{r_B}\sim L$, we expect $Q$ to remain finite. In the limit of $W\to\infty$, $\braket{r_Ar_B}=\braket{r_A}\braket{r_B}$ such that $Q\to 1$. We show results for $Q$ for the disordered spin chain \eqref{eq:H} as a function of $W$ in Fig.~\ref{fig:Q}. The results are qualitatively similar to those of $R$ shown in Fig.~\ref{fig:rarbR}(right).
 
\section*{Details of calculation for MBL state in Fig.~\ref{fig:haarmbl}}

In this section, we sketch the calculation of $\braket{C(r_A,r_B)}$ for the area-law entangled state in Fig.~\ref{fig:haarmbl} constructed by applying a finite-depth, local Haar-random unitary circuit to a product state. For concreteness we will consider $L=4 N$ where $N$ is an integer, however, generalisations to other cases are straightforward. The spatial bipartition implies that sites $\ell= 1,2,\cdots,2N$ lie in subsystem $A$ and sites $\ell = 2N+1,2N+2,\cdots, 4N$ lie in $B$. The structure of the circuit in Fig.~\ref{fig:haarmbl} suggests that one can tripartition the system into a subsystem $A^\prime$ comprising sites $\ell = 1,2,\cdots,2N-2$, a subsystem $A^{\prime\prime}B^{\prime\prime}$ comprising the four sites, $\ell=2N-1,\cdots, 2N+2$, and a subsystem $B^\prime$ with sites $2N+3,2N+4,\cdots,4N$ such that
\eq{
	\ket{\psi_\mathrm{MBL}} = \ket{\psi_{A^\prime}}\otimes\ket{\psi_{A^{\prime\prime}B^{\prime\prime}}}\otimes\ket{\psi_{B^\prime}}\,.
}
We will denote by $\{i_{A^\prime(B^\prime)}\}$ the set of basis states in $A^\prime(B^\prime)$ and by $\{i_{A^{\prime\prime}B^{\prime\prime}}\} = \{i_{A^{\prime\prime}}\}\otimes\{i_{B^{\prime\prime}}\}$ the ones in $A^{\prime\prime}B^{\prime\prime}$. With this notation, the state $\ket{\psi_\mathrm{MBL}}$ can be written as 
\eq{
	\ket{\psi_\mathrm{MBL}} = \sum_{i_{A^\prime}}\sum_{j_{A^\prime}}\sum_{i_{A^{\prime\prime}}i_{B^{\prime\prime}}}\psi_{i_{A^\prime}}\psi_{i_{A^{\prime\prime}}i_{B^{\prime\prime}}}\psi_{i_{B^\prime}}\ket{i_{A^\prime}}\ket{i_{A^{\prime\prime}}}\ket{i_{B^{\prime\prime}}}\ket{i_{B^\prime}}\,.
}
This allows for the correlation $C(r_A,r_B)$ to be expressed as
\eq{
	C(r_A,r_B) = \sum_{r_{A^{\prime\prime}},r_{B^{\prime\prime}}=0}^2 C_{A^\prime}(r_A-r_{A^{\prime\prime}})C_{A^{\prime\prime}B^{\prime\prime}}(r_{A^{\prime\prime}},r_{B^{\prime\prime}})C_{B^\prime}(r_B-r_{B^{\prime\prime}})\,,
}
where 
\eq{
	C_{A^\prime}(r_{A^\prime}=r_A-r_{A^{\prime\prime}}) = \sum_{\substack{i_{A^\prime},j_{A^\prime}:\\r_{i_{A^\prime}j_{A^\prime}}=r_{A^\prime}}}|\psi_{i_{A^\prime}}|^2|\psi_{j_{A^\prime}}|^2\,,
	\label{eq:CAp}
}
is the correlation function in $A^\prime$ (and similarly for $B^\prime$) and 
\eq{
	C_{A^{\prime\prime}B^{\prime\prime}}(r_{A^{\prime\prime}},r_{B^{\prime\prime}}) = \sum_{\substack{\iapp,\japp:r_{\iapp\japp}=r_{A^{\prime\prime}} \\\ibpp,\jbpp: r_{\ibpp\jbpp}=r_{B^{\prime\prime}}}}\psi_{\iapp\ibpp}\psi_{\iapp\jbpp}^\ast\psi_{\japp\ibpp}^\ast\psi_{\japp\jbpp}\,.
}
is the correlation function in $A^{\prime\prime}B^{\prime\prime}$. 

We next evaluate $\braket{C_{A^\prime}(r_{A^\prime})}$ exactly. It will be useful for to define two quantities which are averages of matrix elements of Haar random unitaries as
\eq{
	w = \braket{|u_{\alpha\alpha}|^4}=1/10\,,\quad v=\braket{|u_{\alpha\beta}u_{\alpha\gamma}|^2}=1/20~\quad\mathrm{with}~\beta\neq\gamma\,,
}
where $u$ is a $4\times 4$ Haar random unitary denoted by the grey rectangles in Fig.~\ref{fig:haarmbl}. The structure of the circuit implies that the state $\ket{\psi_{A^\prime}}$ can be expressed as a tensor product over 2-spin states between consecutive odd and even sites, $\ell=2x-1$ and $\ell=2x$ with $x=1,2,\cdots,N-1$. This leads to $\psi_{i_{A^\prime}}=\prod_{x=1}^{N-1}\psi_{i_{A^\prime}}^{(2x-1,2x)}$ from which one can obtain 
$|\psi_{i_{A^\prime}}|^2|\psi_{j_{A^\prime}}|^2 = \prod_{x=1}^{N-1}|\psi_{i_{A^\prime}}^{(2x-1,2x)}|^2|\psi_{j_{A^\prime}}^{(2x-1,2x)}|^2$. Upon averaging over disorder realisations we obtain
\eq{
  \braket{|\psi_{i_{A^\prime}}|^2|\psi_{j_{A^\prime}}|^2} = {w}^{b_0^{(i_{A^\prime}j_{A^\prime})}}{v}^{b_1^{(i_{A^\prime}j_{A^\prime})}+b_2^{(i_{A^\prime}j_{A^\prime})}}\,,
  \label{eq:cAhaar}
}
where $b_\mu^{(i_{A^\prime}j_{A^\prime})}$ is the number of pairs of sites ($2x-1$ and $2x$) where the number of spins different between $i_{A^\prime}$ and $j_{A^\prime}$ is $\mu$. They obviously satisfy the constraints
\eq{
	\sum_{\mu=0}^2 b_\mu^{(i_{A^\prime}j_{A^\prime})} = N-1\,;\quad\mathrm{and}\quad b_1^{(i_{A^\prime}j_{A^\prime})}+2b_2^{(i_{A^\prime}j_{A^\prime})} = r_{i_{A^\prime}j_{A^\prime}}\,.
	\label{eq:constr}
}
We therefore have
\eq{
	\braket{C_{A^\prime}(r_{A^\prime})}=2^{2(N-1)}\sideset{}{'}\sum_{b_0,b_1,b_2}\binom{N-1}{b_0}\binom{N-1-b_0}{b_1}2^{b_1}w^{b_0}v^{b_1+b_2}\,,
	\label{eq:cab0b1b2}
}
where the primed summation denotes that the summation is subject to the constraints in Eq.~\ref{eq:constr} with $r_{i_{A^\prime}j_{A^\prime}}=r_{A^\prime}$. Let us briefly discuss the different factors in the above equation. The first factor is the simply the Hilbert space dimension of subsystem $A^\prime$. Inside the summation, the first and second binomial coefficients are the number of ways of having $b_0$ and $b_1$ pairs with $0$ and $1$ spins different. The factor $2^{b_1}$ accounts for the fact that for pairs with 1 spin different, it could be either the odd or the even site, and finally the factors with $w$ and $v$ account for the state amplitudes. The constraints can be put in explicitly to convert the sum in Eq.~\ref{eq:cab0b1b2} into one just over $b_0$ as 
\eq{
	\braket{C_{A^\prime}(r_{A^\prime})}=2^{2(N-1)}\sum_{b_0 = \max[0,N-1-r_{A^\prime}]}^{\lfloor N-1-\frac{r_{A^\prime}}{2}\rfloor}
  \binom{N-1}{b_0}
  \binom{N-1-b_0}{2(N-1)-2b_0-r_{A^\prime}}
  2^{2(N-1)-2b_0-r_{A^\prime}}
  w^{b_0}v^{N-1-b_0}\,,
  \label{eq:cAsep}
}
which can be evaluated for arbitrarily large systems. An identical calculation yields $\braket{C_{B^\prime}(r_{B^\prime})}$. As far as $C_{A^{\prime\prime}B^{\prime\prime}}(r_{A^{\prime\prime}},r_{B^{\prime\prime}})$ is concerned, it too can be evaluated explicitly as it involves only 4 sites. However, the expression is opaque enough that we do not present it explicitly and evaluate it numerically. Using that and the expression derived in Eq.~\ref{eq:cAsep}, $\braket{C(r_A,r_B)}$ can be evaluated for $\ket{\psi_{\mathrm{MBL}}}$ as in Fig.~\ref{fig:haarmbl}(left) for arbitrarily large system sizes. We use that to obtain the results shown in Fig.~\ref{fig:haarmbl}(centre and right).

\end{document}